\documentclass[aps,preprint,superscriptaddress,pra]{revtex4}
\usepackage[utf8]{inputenc}
\usepackage[english]{babel}
\usepackage{graphicx}
\usepackage{physics}
\usepackage{float}
\usepackage{mathrsfs}
\usepackage{color}
\usepackage{todonotes}
\usepackage{amsmath, amssymb}
\usepackage{placeins}
\usepackage{graphicx}
\usepackage{subcaption}
\usepackage{float}
\usepackage{times}
\usepackage{graphicx,url}
\usepackage[utf8]{inputenc}  
\usepackage{physics}
\usepackage{float}
\usepackage{url}
\usepackage{soul}
\usepackage{graphicx}
\usepackage{hyperref}
\usepackage{dcolumn}
\usepackage{bm}
\usepackage{times}
\usepackage{mathrsfs}
\usepackage{booktabs}
\usepackage{color}
\usepackage{todonotes}

\usepackage[justification=justified]{caption}
\usepackage{ragged2e} 
\begin{document}
\title{Beating the standard quantum limit with single-photon-added coherent states}
\author{Pankaj K. Jha}
\affiliation{Quantum Technology Laboratory $\langle \text{Q}|\text{T}|\text{L}\rangle$, Department of Electrical Engineering and Computer Science, Syracuse University, Syracuse, NY 13244, USA}
\affiliation{Institute for Quantum \& Information Sciences, Syracuse University, Syracuse, NY 13244, USA}
\author{Lakshya Nagpal}
\affiliation{The Institute of Mathematical Sciences (IMSc), C.I.T Campus, Taramani, Chennai 600113, India}
\affiliation{Himalayan Institute for Advanced Study, Gopinath Seva Foundation, Rishikesh 249201, India}
\author{Amir Targholizadeh}
\affiliation{Quantum Technology Laboratory $\langle \text{Q}|\text{T}|\text{L}\rangle$, Department of Electrical Engineering and Computer Science, Syracuse University, Syracuse, NY 13244, USA}
\author{Utkarsh Mishra}
\affiliation{Department of Physics, University of Delhi, Delhi 110007, India}
\author{Konstantin E. Dorfman}
\affiliation{Center for Theoretical Physics and School of Physics and Optoelectronic Engineering, Hainan University, Haikou, China, 570228, China}
\affiliation{Himalayan Institute for Advanced Study, Gopinath Seva Foundation, Rishikesh 249201, India}
\begin{abstract}
The standard quantum limit (SQL), also known as the shot-noise limit, defines how quantum fluctuations of light constrain measurement precision. In a benchmark experiment using the Mach-Zehnder interferometer (MZI), where a coherent state with the average photon number $\langle n\rangle$ is combined with an ordinary vacuum input, the SQL for the phase uncertainty is given by the well-known relation $\Delta\varphi_{\text{SQL}} = 1/\langle n\rangle$. Using a single photon-added coherent state and a weak coherent state as inputs, we report an enhanced phase sensitivity in MZI surpassing the SQL. In stark contrast to other approaches, we focus on the low-photon-number regime, $\langle n\rangle < 10$, and demonstrate that our scheme offers better phase sensitivity compared to the SQL. Beating the SQL at low photon numbers paves the way for the new generation of devices employed in  \textquotedblleft photon-starved\textquotedblright quantum sensing, spectroscopy, and metrology.
\end{abstract}
\date{\today}
\maketitle

Quantum interferometry is the backbone of precision measurements, with applications spanning from gravitational wave detection, quantum sensing, spectroscopy, and optical metrology\cite{scully_quantum_1997,sundar_amplitude-squeezed_1996,pezze_quantum_2018-1,degen_quantum_2017,KDrev}. To implement phase measurements, interferometers based on SU(2) or SU(1,1) symmetries are commonly employed~\cite{yurke_su2_1986,krylov_amplitude-squeezed_1998} . In such setups, the primary objective is to achieve high-precision measurements of specific parameters.
The benchmark experimental platform is based on the Mach-Zehnder interferometer (MZI), which measures phase shifts by utilizing interference between the two optical paths~\cite{helstrom_viii_1976,helstrom_quantum_1969}. The phase sensitivity of an interferometer is fundamentally constrained by the standard quantum limit (SQL), also known as the shot-noise limit, which is determined by the total average photon number of the input states. 
 In particular, the SQL sensitivity of $\varphi$ is given by
\begin{equation}
\Delta\varphi_{\text{SQL}} = 1/{\langle n \rangle}
\end{equation}
where $\langle n\rangle$ is the average number of input photons used to investigate the system~\cite{takeoka_fundamental_2017,ataman_optimal_2019,ataman_phase_2018}.
This intrinsic limitation motivated the exploration of alternative quantum states of light in the attempt to overcome the inherent constraints of classical interferometry, particularly for parameters that cannot be measured directly via conventional techniques~\cite{ataman_phase_2018}.

It was soon realized that the squeezed states of light can enhance the phase sensitivity of an interferometer. The seminal work by Caves demonstrated that squeezed light could improve phase sensitivity beyond the shot-noise limit~\cite{caves_quantum-mechanical_1981,walsh_unlocking_2023,lang_optimal_2014}.
 This technique, proven in both theoretical and experimental contexts, has been utilized in existing gravitational wave detectors, such as LIGO~\cite{smith_path_2009,zhang_noon_2018,oelker_squeezed_2014,caves_quantum-mechanical_1981}. The following experimental demonstrations confirmed the practical applicability of the technique. Subsequent research has focused on optimizing the sensitivity of MZIs with coherent\cite{kitagawa_number-phase_1986} and squeezed vacuum input states~\cite{wei_wigner_2020,takeoka_fundamental_2017,yadav_quantum-enhanced_2024}.

Furthermore, the precision of an interferometric measurement critically depends on the nature of its input light source. Both theoretical and experimental studies have shown that its performance is enhanced with the lowest precision achieved by using thermal light~\cite{kumar_enhanced_2023}, improved by coherent (classical) light, and further improved by non-classical (quantum) light. The latter class of light sources requires a full quantum mechanical description and encompasses states such as single-photon states~\cite{sanders_connection_2014}, squeezed states~\cite{pezze_mach-zehnder_2008,pezze_phase_2007}, twin Fock states~\cite{li_quantum_2024}, Schrödinger's cat states~\cite{wineland_nobel_2013,mishra_non-classicalities_2021,mishra_generation_2020,dakna_generating_1997,dakna_photon-added_1998}, and N00N states~\cite{afek_high-noon_2010,sanders_connection_2014}.

Here, we present a framework showing that a MZI can surpass  SQL when fed with non-classical states. Injecting a single-photon-added coherent state (SPACS)~\cite{PhysRevA.43.492} in one port and a weak coherent state in the other reduces phase uncertainty below the shot-noise level. This  quantum gain, quantified by the normalized phase uncertainty 
$S_{\text{SQL}}(\varphi) = \Delta\varphi / \Delta\varphi_{\text{SQL}}$, arises from the SPACS’s non-classical Wigner-function. These non-classical features interact with the interferometer's unitary transformation to achieve measurement precision surpassing that of classical light sources.

Figure 1 (a) shows a schematic diagram of an MZI featuring two balanced beam splitters, BS$_{1,2}$, and two mirrors, M$_{1,2}$. At the input beam splitter BS$_{1}$, fields propagating in mode $a$ and $b$ gets mixed and the output fields in modes $e$ and $f$ after the output beam splitter BS$_{2}$, are given by~\cite{scully_quantum_1997}
\begin{equation}
\begin{pmatrix}
\hat{f}\\
\hat{e}
\end{pmatrix} =U\begin{pmatrix}
\hat{a}\\
\hat{b}
\end{pmatrix}
\label{eq:MZIsetup}
\end{equation} 
\begin{equation}
U = \begin{pmatrix}
-B & A \\
A & B
\end{pmatrix}
\\
= -e^{i \varphi / 2} \begin{pmatrix}
-\sin \varphi / 2 & \cos \varphi / 2 \\
\cos \varphi / 2 & \sin \varphi / 2
\end{pmatrix}. \label{eq:U_form2}
\end{equation}
All forms satisfy unitarity (\( U U^\dagger = I \)). Here, $(\hat{a}, \hat{b})$, and  ($\hat{e}, \hat{f})$ are the annihilation operators of the input modes ($a,b$) and output modes ($e,f$) respectively and $\varphi$ is the phase shift introduced in the upper arm of the MZI. In this work, we have considered the intensity difference detection scheme at the output of the MZI, where the observable dependent on the phase $\varphi$ is $\hat{O}(\varphi)= \hat{n}_{d}(\varphi) = \hat{e}^{\dagger}(\varphi)\hat{e}(\varphi)-\hat{f}^{\dagger}(\varphi)\hat{f}(\varphi)$. The average value $\langle\hat{n}_{d}(\varphi)\rangle$ and the uncertainty $\langle\left(\Delta \hat{n}_{d}(\varphi)\right)^{2}\rangle$ in the difference between the photon numbers of the two output modes as
\begin{subequations}
\begin{align}
\langle\hat{n}_{d}(\varphi)\rangle&=\left(\langle\hat{n}_{a}\rangle-\langle\hat{n}_{b}\rangle\right) \cos \varphi+2 \text{Re}\{ \langle\hat{a}^{\dagger}\hat{b}\rangle\}\sin \varphi\\
\langle\left(\Delta \hat{n}_{d}(\varphi)\right)^{2}\rangle&=\langle(\Delta \hat{u})^{2}\rangle \cos ^{2}(\varphi)+\langle(\Delta \hat{w})^{2}\rangle \sin ^{2}(\varphi)\nonumber +1/2(\langle\Delta \hat{u} \Delta \hat{w}\rangle+\langle\Delta \hat{w} \Delta \hat{u}\rangle) \sin (2 \varphi) 
\end{align}
\end{subequations}
where $\hat{u}=\hat{a}^{\dagger} \hat{a}-\hat{b}^{\dagger} \hat{b}$ and $\hat{w}=\hat{a}^{\dagger} \hat{b}+\hat{b}^{\dagger} \hat{a}$. Employing the parameter estimation in quantum mechanics, the precision with which a Hermitian operator $\hat{O}(\varphi)$ can be estimated as ~\cite{helstrom_quantum_1969,ataman_quantum_2022}
\begin{equation}
\Delta\varphi = \frac{\sqrt{\langle(\Delta \hat{O}(\varphi))^{2}\rangle}}{\left|\partial\langle\hat{O}(\varphi)\rangle / \partial \varphi\right|}
\end{equation}
Next, we will investigate the possibility of overcoming the SQL limit on the sensitivity of phase measurement by using a photon-added coherent state (PACS) in one of the input modes. 
 
We begin with the simplest case where the input modes $a$ and $b$ are in the ordinary vacuum state $|0\rangle$, and the PACS $|\alpha_{b},m\rangle$ respectively. Introduced by Agarwal and Tara in their seminar paper, PACS is given by~\cite{PhysRevA.43.492}
\begin{equation}
|\alpha_{b}, m\rangle=\frac{\hat{b}^{\dagger m}|\alpha_{b}\rangle}{ (\langle\alpha_{b}|\hat{b}^{m}\hat{b}^{\dagger m}| \alpha_{b}\rangle)^{1/2}}
\end{equation}
where $|\alpha_{b}\rangle$ is a coherent state and $m$ is an integer. The normalization constant $(\langle\alpha_{b}|\hat{b}^{m}\hat{b}^{\dagger m}| \alpha_{b}\rangle)^{1/2} = m !\,L_{m} (-|\alpha_{b}|^{2})$ where $L_{m}(x)$ is the Laguerre polynomial of order $m$. These PACS are intermediate between the Fock and coherent states, exhibiting strong non-classical properties such as quadrature squeezing and sub-Poissonian photon statistics~\cite{PhysRevA.43.492,tara_nonclassical_1993}. Photon addition has been demonstrated through parametric down conversion~\cite{allesandro-2004}, while single-photon sources like colour centers~\cite{jha_nanoscale_2021, akbari-2022} offer a promising alternative pathway to PACS. With an ordinary vacuum state in mode $a$, the second term in Eq. 4(a) and the last term in Eq. 4(b) vanish as $\langle\Delta \hat{u} \Delta \hat{w}\rangle=\langle\Delta \hat{w} \Delta \hat{u}\rangle = 0$. Next, the average value $\langle\hat{n}_{d}(\varphi)\rangle$ and the uncertainty $\langle\left(\Delta \hat{n}_{d}(\varphi)\right)^{2}\rangle$ can be written as
\begin{subequations}
\begin{align}
\langle\hat{n}_{d}\rangle &=-\langle n_{b}\rangle \cos\varphi\\
\langle\left(\Delta \hat{n}_{d}\right)^{2}\rangle &= \langle\left(\Delta \hat{n}_{b}\right)^{2}\rangle\cos^{2}\varphi +\langle n_{b}\rangle\sin^{2}\varphi 
\end{align}
\end{subequations}
where, the average value of the photon number $\langle \hat{n}_{b}\rangle$ and and the number fluctuations $\langle\left(\Delta \hat{n}_{b}\right)^{2}\rangle$ for PACS is given by ~\cite{PhysRevA.43.492}
\begin{subequations}
\begin{align}
\langle \hat{n}_{b}\rangle = &\frac{(m+1)!\,L_{m+1}(-|\alpha_{b}|^{2})}{m!\,L_{m}(-|\alpha_{b}|^{2})}-1\\
\langle\left(\Delta \hat{n}_{b}\right)^{2}\rangle =& \frac{(m+1)(2m+2+|\alpha_{b}|^{2})L_{m+1}(-|\alpha_{b}|^{2})}{L_{m}(-|\alpha_{b}|^{2})}-2(m+1)^2
\end{align}
\end{subequations}
From Eq.~(5,7), we see that in the limit of $\varphi = \pi/2$, the estimated phase sensitivity $\Delta\varphi = 1/\sqrt{\langle n\rangle}$, where $\langle n\rangle = \langle n_{a}\rangle+\langle n_{b}\rangle = \langle n_{b}\rangle$. Thus, we reach the SQL when the inputs are in the ordinary vacuum state $|0\rangle$, and the PACS $|\alpha_{b},m\rangle$. To quantify how much the estimated phase sensitivity $\Delta\varphi$ deviates from the SQL, we define a parameter $\mathcal{S}_{\text{SQL}}(\varphi) = \Delta\varphi/\Delta\varphi_{\text{SQL}}$. Clearly, $\mathcal{S}_{\text{SQL}}(\varphi)<1$ indicates surpassing the SQL, with lower values of $\mathcal{S}_{\text{SQL}}(\varphi)$ indicating improved phase sensitivity. Figure 1 (b,c) shows the plot of $\mathcal{S}_{\text{SQL}}(\varphi)$ against $\varphi$ for different values of $\alpha_{b}$ and $m$, respectively. We observed that $\mathcal{S}_{\text{SQL}}(\varphi)>1$ and it converges to 1 when $\varphi\rightarrow \pi/2$. For a given value of $m$, the convergence $\mathcal{S}_{\text{SQL}}(\varphi) \rightarrow 1$ is faster for a weaker amplitude $\alpha_{b}$ of the PACS. Similarly, for a given amplitude $\alpha_{b}$, we observed a faster convergence $\mathcal{S}_{\text{SQL}}(\varphi) \rightarrow 1$ as $m$ increases. 

Next, we will consider that mode $a$ is in the coherent state $\left|\alpha_{a}\right\rangle$ and the mode $b$ is in the simplest PACS for $m=1$, i.e., SPACS $|\alpha_{b}, 1\rangle$. Using Eq.~4(a) and the relation $\langle\alpha_{a}|\hat{a}^{\dagger} \hat{a}| \alpha_{a}\rangle=|\alpha_{a}|^{2}$, we obtain the difference in photon number as
\begin{equation}
\left\langle\hat{n}_{a}\right\rangle-\left\langle\hat{n}_{b}\right\rangle=\left|\alpha_{a}\right|^{2}-\left(\frac{1+3|\alpha_{b}|^{2}+|\alpha_{b}|^{4}}{1+|\alpha_{b}|^{2}}\right)
\end{equation}
Assuming that the modes are independent of each other, using Eq.~4(a) and using the relation $\langle\alpha_{a}|\hat{a}^{\dagger}| \alpha_{a}\rangle=\alpha_{a}^{*}$, we obtain
\begin{equation}
\langle\hat{a}^{\dagger} \hat{b}\rangle=\alpha_{a}^{*} \alpha_{b}\left(\frac{2+|\alpha_{b}|^{2}}{1+|\alpha_{b}|^{2}}\right)
\end{equation}
From Eqs. (4a, 9, 10) we obtained,
\begin{equation}
\begin{split}
\left\langle\hat{n}_{d}\right\rangle =\left[\left|\alpha_{a}\right|^{2}-\left(\frac{1+3|\alpha_{b}|^{2}+|\alpha_{b}|^{4}}{1+|\alpha_{b}|^{2}}\right)\right]  \cos \varphi+ 2 \operatorname{Re}\left\{\alpha_{a}^{*} \alpha_{b}\right\}\left[\frac{2+|\alpha_{b}|^{2}}{1+|\alpha_{b}|^{2}}\right] \sin \varphi
\end{split}
\end{equation}
Next, we focused on calculating the variance of the difference between the photon numbers, $\langle\left(\Delta \hat{n}_{d}\right)^{2}\rangle$, which has three terms. After lengthy but straightforward calculations, we obtained the first term as 
\begin{equation}
\langle(\Delta \hat{u})^{2}\rangle=\left|\alpha_{a}\right|^{2}+\frac{\left|\alpha_{b}\right|^{2}\left(2+2\left|\alpha_{b}\right|^{2}+\left|\alpha_{b}\right|^{4}\right)}{\left(1+\left|\alpha_{b}\right|^{2}\right)^{2}}
\end{equation}
Similarly, we obtained the second and third terms as
\begin{equation}
\begin{split}
\langle(\Delta \hat{w})^{2}\rangle=\left|\alpha_{a}\right|^{2} +&\frac{\left(1+2\left|\alpha_{a}\right|^{2}\right)\left(1+3\left|\alpha_{b}\right|^{2}+\left|\alpha_{b}\right|^{4}\right)}{1+\left|\alpha_{b}\right|^{2}} +2 \operatorname{Re}\left\{\frac{\alpha_{a}^{* 2} \alpha_{b}^{2}\left(3+\left|\alpha_{b}\right|^{2}\right)}{1+\left|\alpha_{b}\right|^{2}}\right\} \\
&- \left(2 \operatorname{Re}\left\{\frac{\alpha_{a}^{*} \alpha_{b}\left(2+\left|\alpha_{b}\right|^{2}\right)}{1+\left|\alpha_{b}\right|^{2}}\right\}\right)^{2}
\end{split}
\end{equation}
\begin{equation}
\langle\Delta \hat{u} \Delta \hat{w}\rangle +\langle\Delta \hat{w} \Delta \hat{u}\rangle=4\left[\frac{|\alpha_{b}|^{2}}{(1+|\alpha_{b}|^{2})^{2}}\right] \text{Re}\{\alpha_{a}^{*} \alpha_{b}\}
\end{equation}

Figure 2(a, b) shows the density plot of $\mathcal{S}_{\text{SQL}}(\varphi)$ as a function of variables ($|\alpha_{b}|^{2}$, $|\alpha_{a}|^{2}$) and ($\langle n\rangle, \varphi$), respectively. The colored region shows the range of the variables for which $\mathcal{S}_{\text{SQL}}(\varphi)< 1$, i.e., a combination of coherent state $\left|\alpha_{a}\right\rangle$ and SPACS can surpass the SQL. In both, the lowest value of  $\mathcal{S}_{\text{SQL}}(\varphi)$ marginally goes below unity. However, in contrast to previous approaches, this configuration of input states (coherent and SPACS) only requires a few photons to surpass the SQL. Figure 2 (c) shows the plot of $\mathcal{S}_{\text{SQL}}(\varphi)$ against $\langle n\rangle$ for $\alpha_{a}= 1.5 $ and $\varphi = 0$. Here, we observed that for this choice of parameters ($\alpha_{a}, \varphi$), the $\mathcal{S}_{\text{SQL}}(\varphi)$ becomes less than 1 in a particular range of $\langle n\rangle$. For instance, $\mathcal{S}_{\text{SQL}}(\varphi)$ reaches the value 1 at $\langle n\rangle \approx 4.03$ further decreasing and attaining a minimum value of $\sim 0.90$ at $\langle n\rangle \sim 5.2$. Beyond this value of $\langle n\rangle$, $\mathcal{S}_{\text{SQL}}(\varphi)$ begins to increase and eventually goes beyond 1 after $\langle n\rangle \sim$ 8.1. Figure 3 shows the plot of $\mathcal{S}_{\text{SQL}}(\varphi)$ against $\langle n\rangle$ for different values of $m = 1, 2, 3, 4$ and $|\alpha_{a}| = 1.5$. We observed that PACS and a weak coherent state, when used as inputs to an MZI, can surpass the SQL for each case. 

The sub-SQL performance in the SPACS-assisted MZI arises from the nonclassical features of its phase-space representation. The SPACS Wigner function exhibits pronounced negativities~\cite{PhysRevA.43.492, kowalewska-kudlaszyk_wigner-function_2008}, absent in classical states such as coherent or thermal fields. When injected into an MZI, this nonclassicality interacts with the interferometer’s phase-dependent unitary transformation, reshaping the Wigner distribution and generating interference fringes linked to enhanced Fisher information and phase sensitivity~\cite{fan_application_1987, wei_wigner_2020}. Evaluating the output Wigner function for various internal phases $\varphi$ reveals how these phase-space features govern the observed enhancement. 

The Wigner function is a quasi-probability distribution that provides a comprehensive phase-space representation of the quantum state~\cite{fan_application_1987}. The Wigner function is given by \begin{equation} W(\alpha_a, \alpha_b^*) = \frac{2}{\pi} \int_{-\infty}^\infty \langle \alpha_a + \alpha_b | \rho | \alpha_a - \alpha_b \rangle e^{-2i \, \text{Im}(\alpha_a \alpha_b^*)} \, d^2\alpha_b, \end{equation} where \(\rho\) is the density operator of the quantum state, and \(\alpha_a\) and \(\alpha_b\) are complex variables representing the phase-space coordinates. The relation between the output Wigner function and the input Wigner function can be written as ~\cite{dowling_wigner_1994} 
\begin{equation} 
\begin{split} W_o\left(\alpha_{a,i}, \alpha_{b,i}\right) = W_a\left(\alpha_{b,i} A^* - \alpha_{a,i} B^*\right) W_b\left(\alpha_{a,i} A^* + \alpha_{b,i} B^*\right) 
\end{split} 
\end{equation} 
For the case when one input is a coherent state and the other is SPACS, the Wigner function of the outgoing field can be obtained 
\begin{equation} 
\begin{split}  W(\alpha_a, \alpha_b) = \frac{4 N_m^2 (-1)^{m+n} m!}{\pi^2} e^{-2(|\alpha_a - \alpha_{a,0}|^2 + |\alpha_b - \alpha_{b,0}|^2)}  L_m\left(2|\alpha_a - \alpha_{a,0}|^2\right) L_n\left(2|\alpha_b - \alpha_{b,0}|^2\right). 
\end{split} 
\label{eq:wigtot} 
\end{equation}

Here \(L_{m}(x)\) is the Laguerre polynomial. Equation~\ref{eq:wigtot} gives the total Wigner function of the output radiation/field. The negative value of the Wigner function characterizes the non-classicality of the outgoing field. Figure~\ref{fig:SPACS_Wigner} presents the Wigner function for different MZI configurations. In Fig.~4(a), the combined input ($m=1$) shows interference-induced negativity at the origin~\cite{Arman:21}, a hallmark of photon addition. Figures~4(b–c) depict the output Wigner function for varying phases $\varphi$, where negativity decreases as $\varphi$ increases, reflecting phase-dependent redistribution~\cite{wei_wigner_2020}. For higher-order PACS ($m=4$), Figs.~4(d–f) display more intricate interference fringes at $\varphi=0$ that become smoother at $\varphi=3.77$, consistent with earlier analyses of non-Gaussian phase evolution~\cite{fan_application_1987, shukla_broadening_2021}. At small $\varphi$, pronounced negativities near the phase-space origin correspond to higher Fisher information and sub-SQL sensitivity~\cite{wei_wigner_2020}. As $\varphi$ increases, these features diminish, driving the state toward classicality~\cite{shukla_broadening_2021}. This confirms that Wigner-function negativity reliably signals quantum advantage, directly linking phase-space nonclassicality to enhanced metrological precision.

In summary, we investigated the precision with which the phase shift $\varphi$ between the two arms of an MZI can be measured when one of the input modes $b$ of the MZI is in the PACS ($|\alpha_{b}, m\rangle$). We show that if the other input mode $a$ is in the ordinary vacuum state $|0\rangle$, then this combination of input states \textit{does not} break the SQL. However, when the mode $a$ is injected with a weak coherent state ($|\alpha_{a}\rangle$), and mode is in the simplest PACS ($m$=1) i.e. SPACS ($|\alpha_{b}, 1\rangle$), this combination of input states \textit{beats} the SQL for low photon numbers. We analyze the sub-SQL results using the Wigner function, which highlights the subtle non-classical behavior of SPACS in the low-photon-number regime.   

\clearpage
\begin{figure}[t]
 \centerline{{\includegraphics[width=\textwidth]{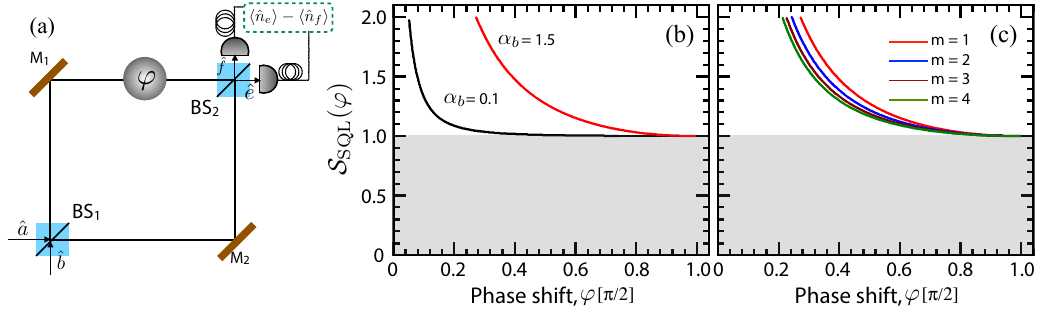}}}
  \caption{\justifying (a) Schematic diagram of an MZI with input modes ($a,b$) and output modes ($e,f$). A phase shift ($\varphi$) is introduced in the upper arm relative to the lower arm and an intensity difference detection scheme ($\langle{n}_{e}\rangle-\langle{n}_{f}\rangle$) at the output was employed to extract the phase shift $\varphi$. (b) Plot of the normalized phase uncertainty $\mathcal{S}_{\text{SQL}}(\varphi)=\Delta\varphi/\Delta\varphi_{\text{SQL}}$ as a function of phase shift ($\varphi$) for $\alpha_{b}$= 0.1 (black line) and $\alpha_{b}$= 1.5 (red line). We see that for both cases $\mathcal{S}_{\text{SQL}}(\varphi) > 1$ i.e. SPACS cannot beat the SQL and $\mathcal{S}_{\text{SQL}}(\varphi)  \rightarrow 1$ for $\varphi\rightarrow \pi/2$. (c)  Plot of $\mathcal{S}_{\text{SQL}}(\varphi)$ as a function of phase shift ($\varphi$) for $\alpha_{b}$= 1.5 and $m = 1, 2, 3, 4, 5$.}
 \end{figure}
 
 \clearpage
 \begin{figure}[t]
 \centerline{\includegraphics[width=\textwidth]{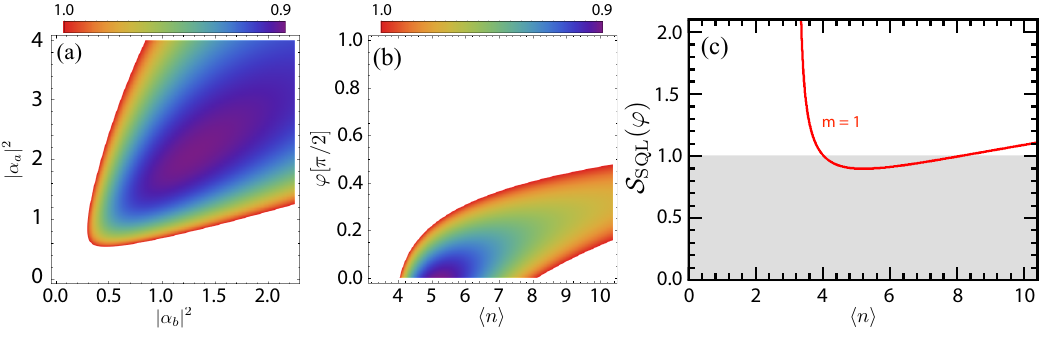}}
  \caption{\justifying Density plot of $\mathcal{S}_{\text{SQL}}(\varphi)$ as a function of (a) $|\alpha_{a}|^{2}$ and $|\alpha_{b}|^{2}$ for $\varphi = 0$ and (b) phase shift ($\varphi$) and average photon number $\langle n\rangle$ for $|\alpha_{a}|$= 1.5. The colored region shows the range of the parameters in (a,b) for which $\mathcal{S}_{\text{SQL}}(\varphi) < 1$ i.e. the coherent state $\left|\alpha_{a}\right\rangle$ and the SPACS $|\alpha_{b}, 1\rangle$ as inputs can beat the SQL. (c) Plot of $\mathcal{S}_{\text{SQL}}(\varphi)$ as a function of the average photon number $\langle n\rangle$ for $|\alpha_{a}|$= 1.5 and $m = 1$. We see that there is small range of $\langle n\rangle \sim$ $ 4.03 \leq \langle n\rangle  \leq 8.1$ where $\mathcal{S}_{\text{SQL}}(\varphi) < 1$.}
 \end{figure}
 
 \clearpage
 \begin{figure}[b] 
\centerline{{\includegraphics[scale = 1.1]{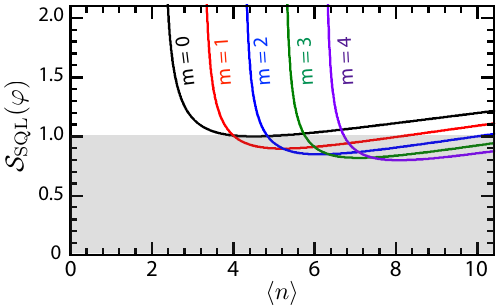}}} 
\caption{\justifying Plot of of $\mathcal{S}_{\text{SQL}}(\varphi)$ as a function of average photon number $\langle n\rangle$ for $|\alpha_{a}|$= 1.5, $\varphi = 0$ for different PACS $|\alpha_{b},m\rangle$ ($m$ = 0, 1, 2, 3, 4). Here, $m=0$ corresponds to no photon addition, i.e., the input modes are in coherent states.} 
\end{figure} 

\clearpage
\begin{figure}[htbp]
\centering
\includegraphics[width=\textwidth]{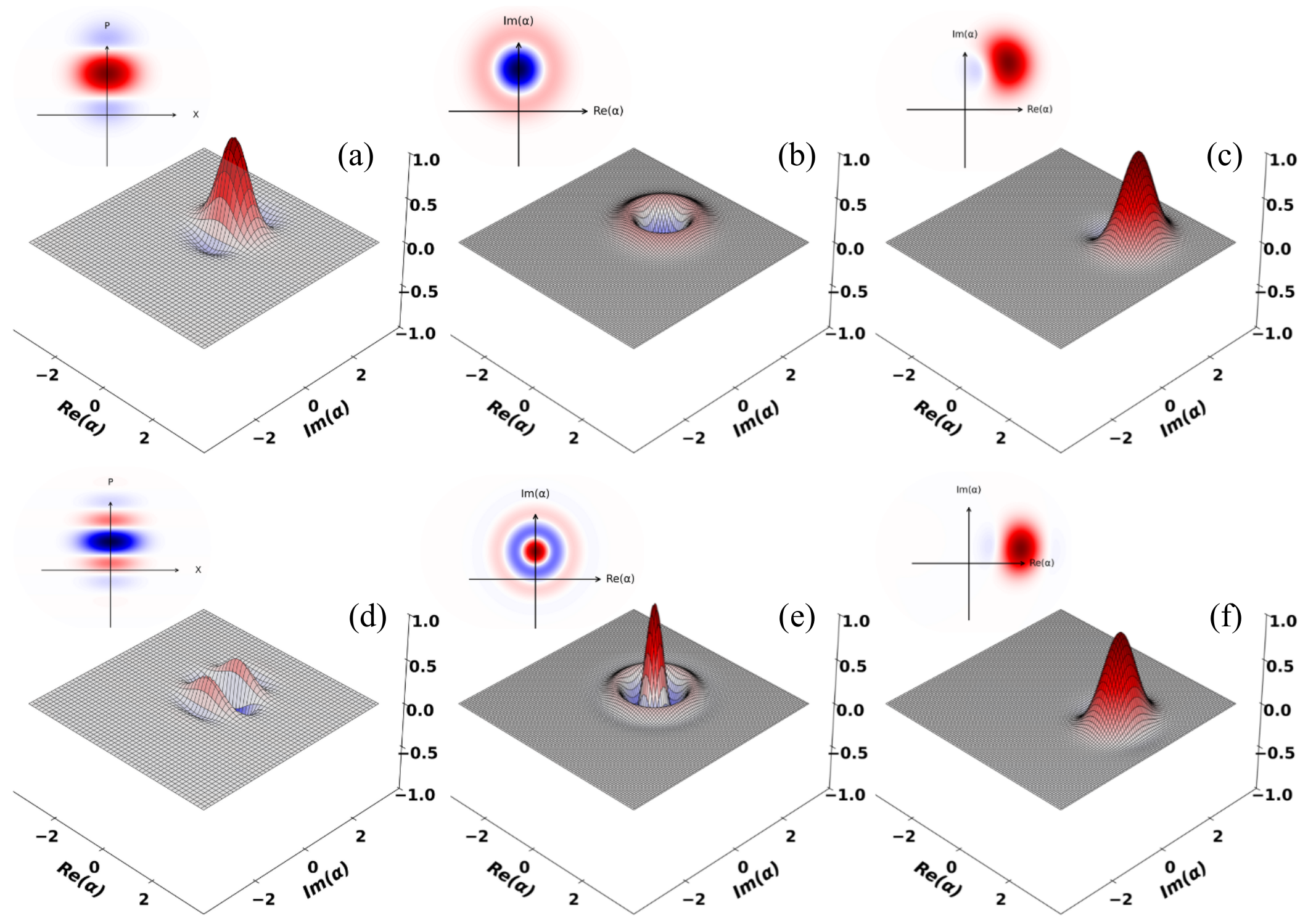}
\caption{ \justifying Phase-space Wigner function profiles for SPACS combined with a coherent state in an MZI. The vertical axis denotes the computed Wigner function value, highlighting the degree of nonclassicality. (a) Combined Wigner function of a SPACS ($m=1$ where $\alpha_a = 1.17 $ and $\alpha_b = 1.5 $) and a coherent state at the MZI input, showing interference-induced negativity at the origin. (b) Single-mode SPACS Wigner function after propagation through the MZI for $m=1$ and $\varphi=0$, preserving pronounced nonclassical features. (c) Output Wigner function for $m=1$ and $\varphi=5.03$, where the negativity diminishes and the profile becomes more Gaussian-like. (d) Input Wigner function for a higher-order SPACS ($m=4$ where $\alpha_a = 0.819 $ and $\alpha_b = 1.5 $), exhibiting complex interference fringes. (e) Wigner function at the MZI output for $m=4$ and $\varphi=0.00$, with sharp central negativity retained. (f) Output Wigner function for $m=1$ and $\varphi=3.77$, where nonclassical features are suppressed due to phase-induced redistribution. }
\label{fig:SPACS_Wigner}
\end{figure}

\clearpage

\bibliography{manuscript}
\end{document}